\documentclass{elsart}
\usepackage{amssymb}
\usepackage{amsmath}
\usepackage{graphics}
\usepackage{epsfig}
\usepackage{amssymb}
\usepackage{subfigure}
\input epsf
\newcommand{\bmath}[1]{\mbox{\boldmath ${#1}$}}
\newcommand{\dd}{\mbox{\rm d}}
\newcommand{\dpce}{\ensuremath{\vec{d}p\to (pp)n}}
\newcommand{\dpnce}{\ensuremath{\vec{d}p\to (pn)p}}
\newcommand{\npdpce}{\ensuremath{dp\to (pp)n}}
\def\fmn#1#2{\mbox{${\textstyle \frac{#1}{#2}}$}}
\newcommand{\ww}{\mbox{$\,$}}

\begin{document}
\vspace*{-1cm}
\begin{frontmatter}
\title{Vector and tensor analysing powers in deuteron--proton
breakup reactions at intermediate energies}
\author[ikp,tbilisi]{D.~Chiladze}%
\footnote{Corresponding author: d.chiladze@fz-juelich.de},
\author[Grenoble]{J.~Carbonell},
\author[dubna]{S.~Dymov},
\author[dubna2]{V.~Glagolev},
\author[ikp]{M.~Hartmann},
\author[ikp]{V.~Hejny},
\author[erlangen,tbilisi]{A.~Kacharava},
\author[ikp,tbilisi]{I.~Keshelashvili},
\author[munster]{A.~Khoukaz},
\author[ikp]{H.R.~Koch},
\author[dubna]{V.~Komarov},
\author[ikp,cracow]{P.~Kulessa},
\author[dubna]{A.~Kulikov},
\author[dubna,tbilisi]{G.~Macharashvili},
\author[cologne]{Y. Maeda},
\author[munster]{T.~Mersmann},
\author[ikp,dubna]{S.~Merzliakov},
\author[gatchina]{S.~Mikirtytchiants},
\author[ikp]{A.~Mussgiller},
\author[tbilisi]{M.~Nioradze},
\author[ikp]{H.~Ohm},
\author[ikp]{F.~Rathmann},
\author[ikp]{R.~Schleichert},
\author[ikp]{H.J.~Stein},
\author[ikp]{H.~Str\"oher},
\author[dubna]{Yu.~Uzikov},
\author[erlangen,dubna]{S.~Yaschenko},
and
\author[london]{C.~Wilkin}.
%
%
\address[ikp]{Institut f\"ur Kernphysik, Forschungszentrum J\"ulich,
52425 J\"ulich, Germany}
\address[tbilisi]{High Energy Physics Institute, Tbilisi State
University, 0186 Tbilisi, Georgia}
\address[Grenoble]{ Laboratoire de Physique Subatomique et de Cosmologie, 38026
Grenoble, France}
\address[dubna]{Laboratory of Nuclear Problems, JINR, 141980 Dubna, Russia}
\address[dubna2]{Laboratory of High Energies, JINR, 141980 Dubna, Russia}
\address[erlangen]{Physikalisches Institut II, Universit\"at
Erlangen--N\"urnberg, 91058 Erlangen, Germany}
\address[munster]{Institut f\"ur Kernphysik, Universit\"at M\"unster,
48149 M\"unster, Germany}
\address[cracow]{Institute of Nuclear Physics, 31342 Cracow, Poland}
\address[cologne]{Institut f\"ur Kernphysik, Universit\"at zu K\"oln, 50937 K\"oln,
Germany}
\address[gatchina]{High Energy Physics Department, PNPI, 188350 Gatchina, Russia}
\address[london]{Physics and Astronomy Department, UCL, London, WC1E 6BT, UK}
%
%
\abstract{Vector and tensor analysing powers of the \dpce\
(charge--exchange) and \dpnce\ (non--charge--exchange) breakup
reactions have been measured with the ANKE spectrometer at the
COSY ring at a deuteron beam energy of 1170{\ww}MeV for small
momentum transfers to the low excitation energy $(pp)$ or $(pn)$
systems. A quantitative understanding of the values of $A_{xx}$
and $A_{yy}$ for the charge--exchange reaction is provided by
impulse approximation calculations. The data suggest that
spin--flip isospin--flip transitions, which dominate the
charge--exchange breakup of the deuteron, are also important in
the non--charge--exchange reaction.}}
%
%
\end{frontmatter}
%
%

It was suggested several years ago that the charge--exchange
breakup of medium energy deuterons should show significant
polarisation effects~\cite{BW1,BW2}. The tensor analysing powers
of the \dpce\ reaction were predicted to be especially large for
excitation energies $E_{pp}$ of the final proton--proton pair
below about 3${\ww}$MeV, provided that the momentum transfer $q$
from the incident deuteron to the di-proton also remained small.
Under these conditions, the final--state interaction in the
$^1\!S_0$ state of the $pp$ system is very strong. The impulse
approximation predictions were tested successfully for very small
$E_{pp}$ at beam energies of $T_d=1.6\ww{}$GeV and 2.0\ww{}GeV
using both hydrogen and deuterium
targets~\cite{Ellegaard1,Ellegaard2,Sams95a}. A refined numerical
evaluation~\cite{Carbonell} proved that the theoretical
description was equally valid over a larger $E_{pp}$ range at 200
and 350{\ww}MeV~\cite{Kox}. The cross section and tensor analysing
powers are so large that the reaction could form the basis of a
deuteron polarimeter with a high figure of merit~\cite{POLDER},
which was used to measure the polarisation of the recoil deuteron
in large momentum transfer elastic electron--deuteron scattering
at JLab~\cite{JLab}.

One major feature of the reaction is that the differential cross
section and the deuteron Cartesian analysing powers $A_{xx}$ and
$A_{yy}$ are directly related to the magnitudes of the spin--spin
neutron--proton charge--exchange amplitudes~\cite{BW2}. These
govern the spin--transfer observables in the $\vec{n}p\to
\vec{p}n$ reaction at small momentum transfers from neutron to
proton. This raises the possibility of using the deuteron
charge--exchange reaction to obtain information about $pn$
observables above say 800\ww{}MeV where, despite much work carried
out at Saclay~\cite{Ball}, accurate polarisation data are more
sparse~\cite{SAID}. To lend credence to such an approach, it is
necessary to make detailed \dpce\ measurements in a domain where
one can have confidence in the nucleon--nucleon amplitudes used in
the analysis. We therefore report here on the study of $A_y$,
$A_{xx}$ and $A_{yy}$ at $T_d=1170{\ww}$MeV ($p_d=2400{\ww}$MeV/c)
up to $q=130\ww{}$MeV/c. Quantitative agreement with predictions
based upon an up--to--date phase shift analysis~\cite{SAID,Igor}
is obtained, and this will allow the programme of using the
charge--exchange reaction for $pn$ studies to go ahead at
COSY~\cite{SPIN}.

Though the situation regarding the \dpce\ reaction looks fairly
clear, that of the non--charge--exchange \dpnce\ breakup is far
more complex. Even in the $S$--wave one has to consider the
production of both $^1\!S_0$ and $^3\!S_1$ $(pn)$ states and in
general the number of relevant low energy $pn$ final states is
twice that of $pp$, as is the number of elastic $NN$ amplitudes
that provide the driving force in the process. In a first
exploration, we also took data in parallel on the \dpnce\ reaction
in the range $70<q<200\ww{}$MeV/c, where the excitation energy
$E_{pn}$ in the final $pn$ system was similarly constrained to be
at most a few MeV. The results show some similarities to those of
\dpce\ under similar kinematic conditions and reinforce the belief
that, at low momentum transfers, this reaction is also dominated
by spin--flip isospin--flip transitions.

The experiment was carried out at the COSY COoler SYnchrotron of
the Forschungszentrum J\"ulich using the ANKE magnetic
spectrometer, located at an internal target position of the
storage ring. Although ANKE has several detection
possibilities~\cite{Barsov01}, only those of the Forward Detector
(FD) system were used here to detect the two fast protons from the
\dpce\ reaction~\cite{Chiladze0}. The FD consists of multiwire
chambers for track reconstruction and three layers of a
scintillation hodoscope that permit time--of--flight and
energy--loss determinations~\cite{Dymov}. The tracking system
gives a momentum resolution of better than 1\%. While fast protons
from the $dp\to (pn)p$ reaction were also measured in the FD, the
slow recoil protons in the energy range $2.5<T_p<32\ww{}$MeV were
detected in a Silicon Tracking Telescope (STT) placed inside the
target chamber~\cite{Chiladze0,Inti,Andreas}. This provided a
polar and azimuthal angular coverage of
 $75.6^{\circ}<\theta_{\rm STT}<116.4^{\circ}$ and
 $-21.8^{\circ}<\phi_{\rm STT}<19.4^{\circ}$ respectively.

%
\begin{figure}[htb]
\begin{center}
\centerline{\epsfxsize=9cm {\epsfbox{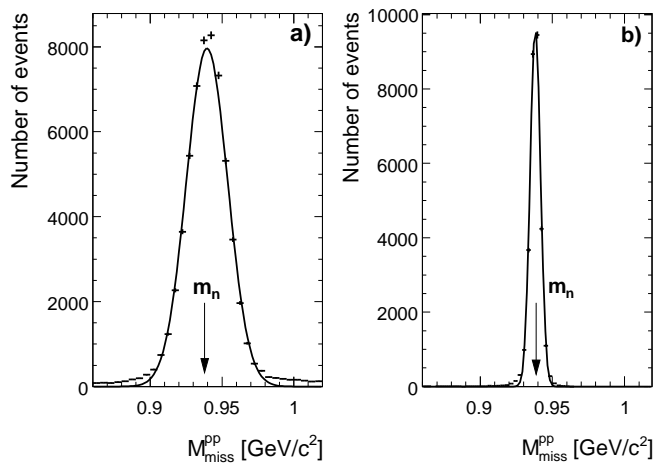}}}
\end{center}
\vspace{-5mm} \caption{\label{pi0} Missing masses of the
$p(d,2p)X$ reaction obtained by measuring a) both final fast
protons in the FD, and b) a fast proton in the FD and a slow one
in the STT. Though there is slightly more background in case a),
and the Gaussian peak is much wider ($\sigma = 13\ww{}$MeV/c$^2$
compared to $4\ww{}$MeV/c$^2$), there is no difficulty in
extracting the $p(d,2p)n$ signal in either case. Events were
retained in $\pm2.5\,\sigma$ regions around the peaks.}
\end{figure}
%

In the deuteron charge--exchange reaction, two fast protons are
emitted in a narrow forward cone with momenta around half that of
the deuteron beam. As described in Ref.~\cite{Chiladze0}, such
coincident pairs could be clearly identified using information
from the FD system, the coverage in the laboratory polar angle
being between $0^{\circ}$ and $6^{\circ}$. Because of this limited
acceptance, there are strong kinematic correlations between the
magnitudes of the two momenta for two-- and three--body final
states. For charge--exchange candidates selected in this way, the
times of flight for the two particles from the target to the
hodoscope were calculated, assuming the particles to be protons.
The difference in these times of flight could be compared with the
measured time difference for those events where the particles hit
different counters in the hodoscope. This selection, which rejects
about 20\% of the events, essentially eliminates background, for
example from $dp$ pairs associated with $\pi^0$ production. The
resulting missing--mass distribution for identified $ppX$ events
shows a clean neutron peak in Fig.~\ref{pi0}a at
$M_X=940.4\pm0.2{\ww}$MeV/c$^2$ ($\sigma = 13\ww{}$MeV/c$^2$),
sitting on top of a slowly varying 2\% background.

The non--charge--exchange breakup $dp\to (pn)p$ reaction was
isolated by first identifying slow protons emerging from the
target in the STT and then looking at the momenta of charged
particles detected in coincidence in the FD. There is a gap of at
least 800\ww{}MeV/c between high momentum, elastically scattered,
deuterons and protons from the $dp\to (pn)p$ reaction, which have
about half the beam momentum. The angular and energy resolution in
the STT leads to a good excitation energy determination of the
fast $pn$ pair and, as a result, the missing--mass resolution on
the neutron peak in Fig.~\ref{pi0}b is better than in \ref{pi0}a,
with essentially no background.

The COSY polarised ion source was set up to provide a sequence of
an unpolarised state, followed by seven combinations of deuteron
vector ($P_z$) and tensor ($P_{zz}$) polarisations, where $z$ is
the quantisation axis in the source frame of reference. The
determination of the actual polarisations of the beam, through the
measurement of a variety of nuclear reactions, is detailed in
Ref.~\cite{Chiladze0}. It is shown there that the values of $P_z$
were on average about 74\% of the ideal figures that could be
obtained from the source, whereas the corresponding reduction
factor for $P_{zz}$ was typically around 59\%.

Having identified the \npdpce\ events, these were binned in
intervals of di-proton excitation energy $E_{pp}$ and
three--momentum transfer $q=\sqrt{-t}$, and corrected for
luminosity with the help of the beam current information in order
to evaluate the analysing powers. In the right--handed coordinate
system of the reaction frame, the beam defines the $z$--direction
while the stable spin axis of the beam points along the
$y$--direction, which is perpendicular to the COSY orbit. The
numbers $N(q,\phi)$ of di-protons produced at momentum transfer
$q$ and azimuthal angle $\phi$ with respect to the $x$--direction
are given in terms of the beam polarisations by
\begin{equation}
N(q,\phi)=N_0(q)\,\left[1+\fmn{3}{2} P_{z}A_{y}(q)\cos\phi
+\fmn{1}{2}P_{zz}\!\left\{A_{yy}(q)\cos^2\phi
+A_{xx}(q)\sin^2\phi\right\}\right]\!, \label{eqn_fit}
\end{equation}
where $N_0(q)$ are the numbers for an unpolarised beam, and $A_y$
($A_{yy}$, $A_{xx}$) are vector (tensor) analysing powers of the
\dpce\ reaction~\cite{Ohlsen}.

%
\begin{figure}[hbt]
\begin{center}
{\epsfig{file=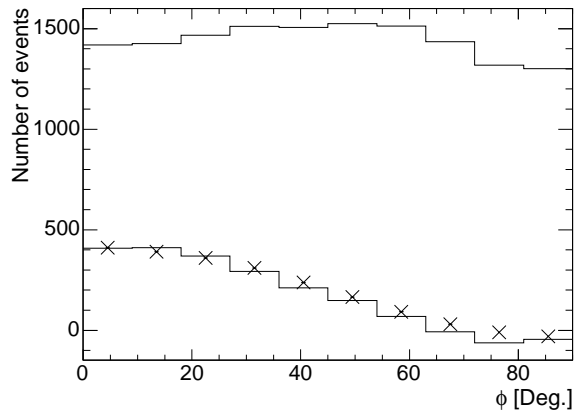,height=6cm}}%
\caption{$\phi$--distribution for the combination of counts
$(N_3-N_2)$ (lower histogram) and $(N_3+N_2)/2$ (upper) for spin
states 2 and 3 for $E_{pp}<1{\ww}$MeV and $40 < q < 60{\ww}$MeV/c
of the di-protons from the $dp\to (pp)n$ reaction at 1170{\ww}MeV.
The crosses denote the calculated values of $(N_3-N_2)$ from
Eq.~(\ref{eqn_fit}) using the known values of $P_{zz}$ in states 2
and 3, the extracted tensor analysing powers, and the values of
$N_0=(N_3+N_2)/2$ for the different $\phi$ bins.\label{phi23}}
\end{center}
\end{figure}
%

Measurements were made using the eight available spin states of
the source. For each of the seven intervals in $q$ (about
20\ww{}MeV/c width), the yield was sorted in 120 bins of
$3^{\circ}$ width in $\phi$, corresponding roughly to the
resolution of the spectrometer. Using the known values of the beam
polarisations $P_z$ and $P_{zz}$~\cite{Chiladze0}, together with
the average values of the trigonometric functions within each
$\phi$--bin, it is possible to determine the three observables
$A_y$, $A_{xx}$, and $A_{yy}$ in a fit of Eq.~(\ref{eqn_fit}). For
each of the seven $q$--intervals, the resulting information from
the eight different spin states was combined using a weighted
average.

The effect of the beam tensor polarisation can be seen immediately
just by comparing spin states 2 and 3, where the ideal
polarisations are $\left(P_z,\,P_{zz}\right)= \left(1/3,1\right)$
and $\left(-1/3,-1\right)$, respectively. The
$\phi$--distributions of two combinations of the number of events
for these states are shown in Fig.~\ref{phi23} for $E_{pp}<
1{\ww}$MeV and $40< q< 60{\ww}$MeV/c. When events from all four
quadrants, at $\phi$, $180^{\circ}-\phi$, $180^{\circ}+\phi$, and
$360^{\circ}-\phi$, are considered together, the term depending on
$A_y$ is largely cancelled. In the ideal case, the average of the
counts $(N_3+N_2)/2$ corresponds to that of an unpolarised beam,
$N_0$, with a $\phi$--dependence that reflects the acceptance of
the spectrometer, independent of the beam polarisation. The
difference $(N_3-N_2)$ is strongly modulated by polarisation
effects. The crosses in Fig.~\ref{phi23} show the calculated
values of $(N_3-N_2)$ using the known values of $P_{zz}$ for the
spin states 2 and 3, the extracted tensor analysing powers, and
the measured values of $(N_3+N_2)/2$.

\begin{figure}[hbt]
\begin{center}
\subfigure{\epsfig{file=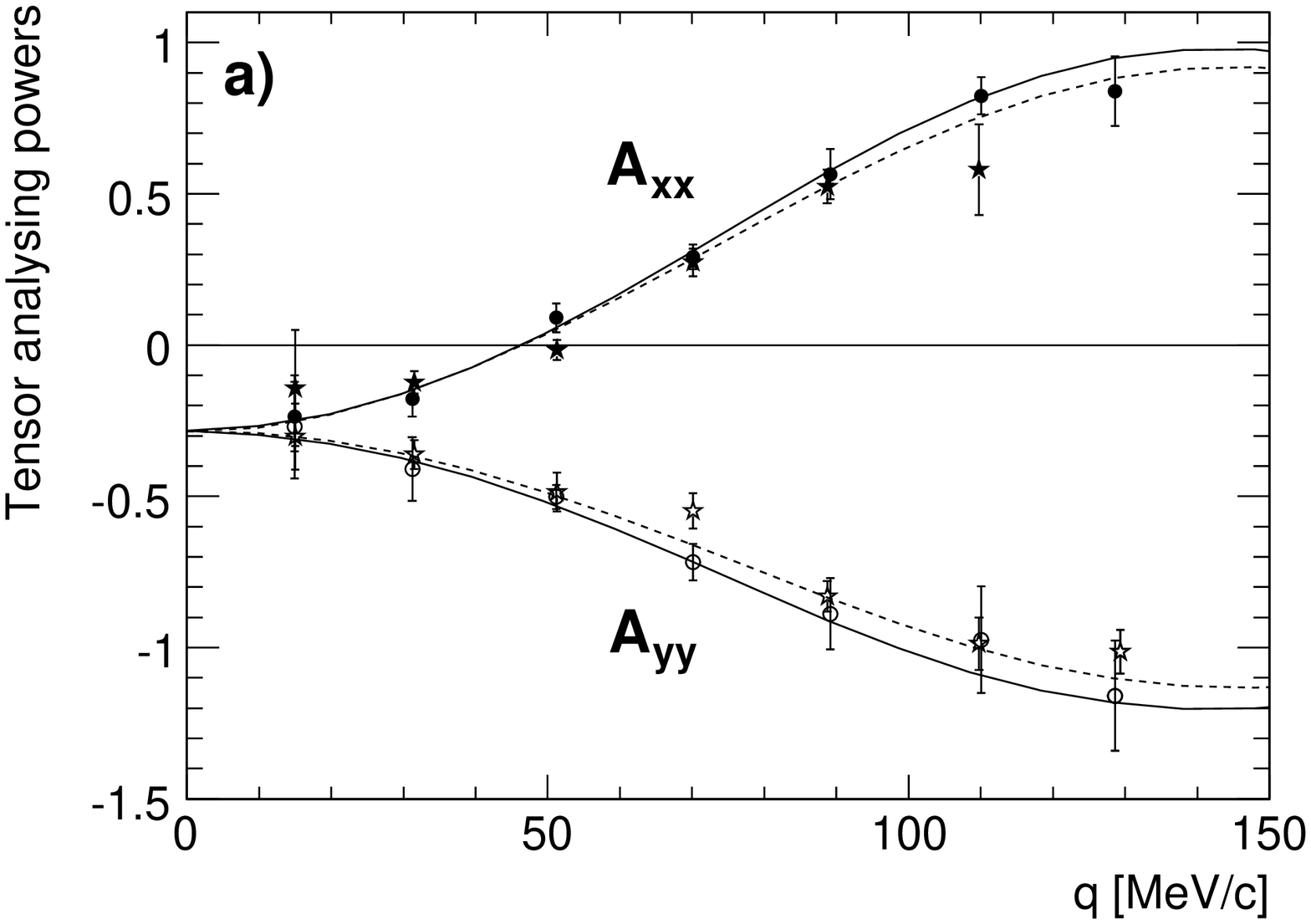,height=5cm}}%
\subfigure{\epsfig{file=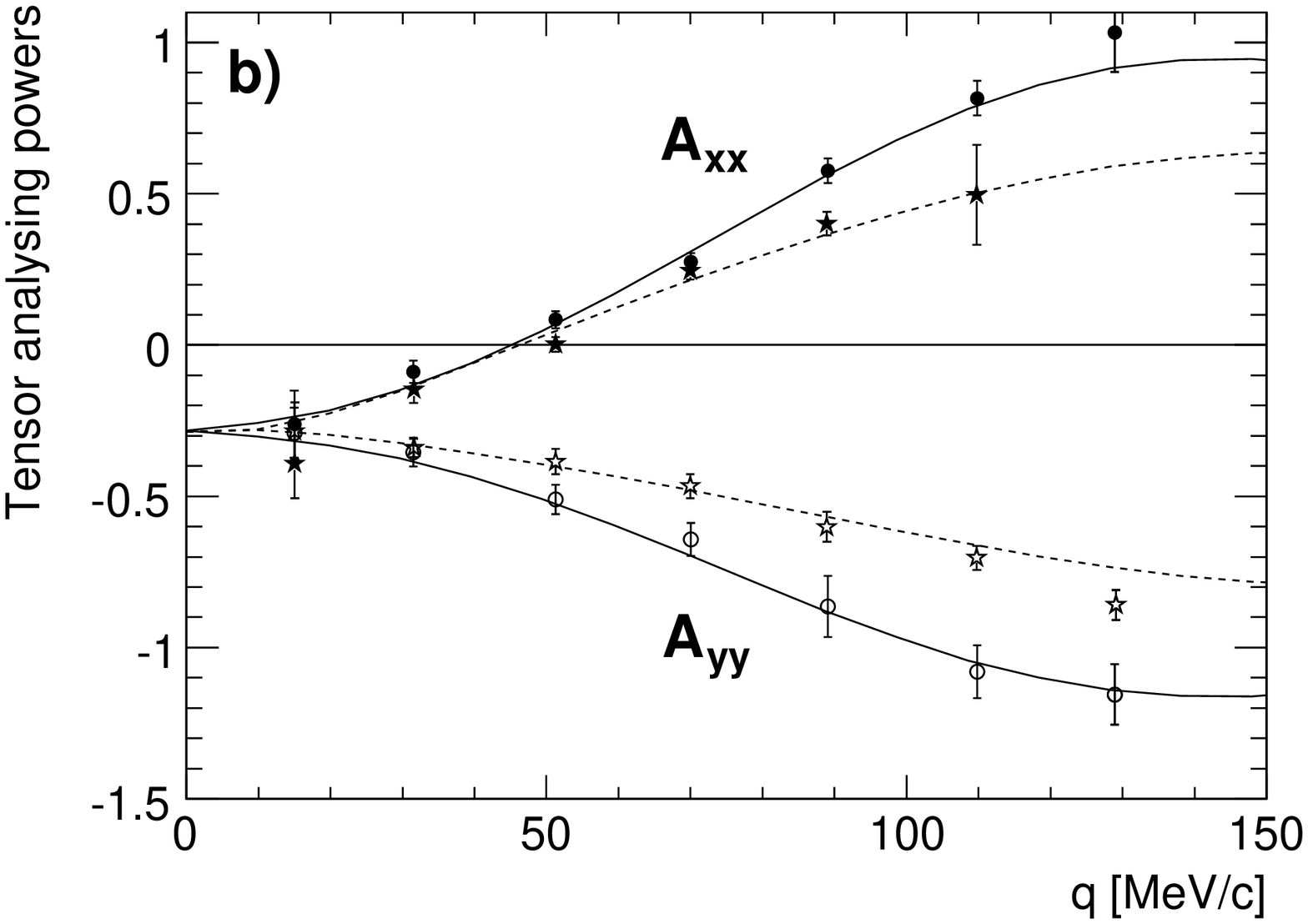,height=5cm}}%
\caption{\label{Axx-Ayy}Cartesian tensor analysing powers $A_{yy}$
(open symbols) and $A_{xx}$ (closed) of the $\vec{d}p\to(pp)n$
reaction for a) $0.1<E_{pp}<1\ww{}$MeV, and b)
$1<E_{pp}<3\ww{}$MeV. The circles correspond to events where
$|\cos\theta_{qk}|<0.5$ whereas the stars denote
$|\cos\theta_{qk}|>0.5$. The solid and broken curves, which
involve respectively the same angular selection, follow from the
impulse approximation program of Ref.~\cite{Carbonell}, for which
the 585{\ww}MeV input amplitudes were taken from
Ref.~\cite{SAID,Igor}. The error bars include the uncertainties in
the beam polarisation of about 4\%~\cite{Chiladze0}.}
\end{center}
\end{figure}

In impulse approximation the vector analysing power is predicted
to vanish for small $E_{pp}$~\cite{BW2} and our results are
consistent with this. The averages over the whole range in $q$ are
$\langle A_y\rangle\ = -0.001\pm 0.005$ and $\langle A_y\rangle\ =
-0.004 \pm 0.004$ for $0.1<E_{pp}<1{\ww}$MeV and
$1<E_{pp}<3{\ww}$MeV, respectively. The 0.1\ww{}MeV lower limit
arises from the requirement that the protons hit different
counters of the FD. In order to improve the precision for the
comparison with theory, we then imposed $A_y=0$ and extracted
values for the two tensor analysing powers. In this way the error
bars are reduced compared to the fit with unconstrained $A_y$, in
particular for the highest $q$ bins. The results are shown in
Fig.~\ref{Axx-Ayy} as functions of $q$, separately for the two
$E_{pp}$ bins. The resolution in $E_{pp}$ is about 0.3\ww{}MeV at
3\ww{}MeV and better at lower $E_{pp}$~\cite{komarov}. Due to the
limited ANKE angular coverage, the acceptance gets steadily poorer
as $q$ and $E_{pp}$ increase, so that values of $A_{yy}$ could
only be determined for $q< 130{\ww}$MeV/c, with a slightly lower
limit in the case of $A_{xx}$. As explained later, the data are
further divided into two groups depending on the angle
$\theta_{qk}$ between $\vec{q}$ and the final $pp$ relative
momentum $\vec{k}$.

%
%

\begin{figure}[htb]
\begin{center}
\centerline{\epsfxsize=8cm{\epsfbox{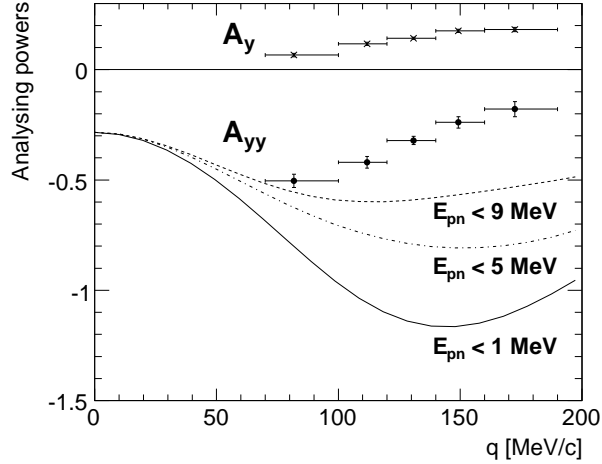}}}
\end{center}
\vspace{-1cm} \caption{\label{Ayynp} Cartesian analysing powers
$A_y$ (crosses) and $A_{yy}$ (circles) for the ${d}p\to (pn)p$
reaction for $E_{pn}<5${\ww}MeV in 20\ww{}MeV/c or 30\ww{}MeV/c
bins in momentum transfer. The curves correspond to the
predictions of $A_{yy}$ from the charge--exchange impulse
approximation program of Ref.~\cite{Carbonell}, as used for
Fig.~\ref{Axx-Ayy}; $E_{pn}< 1{\ww}$MeV (solid),
$E_{pn}<5{\ww}$MeV (chain), and $E_{pn}<9{\ww}$MeV (dashed).}
\end{figure}

Turning to the \dpnce\ reaction, the positioning of a relatively
small aperture telescope in the horizontal plane means that only
$A_y$ and $A_{yy}$ could be measured here. Furthermore, since
$\langle\cos^2\phi\rangle\approx 0.96$, there is a small
contamination in $A_{yy}$ from the $A_{xx}$ contribution. In the
extreme case that this is large and opposite in sign, as found for
the charge exchange reaction in Fig.~\ref{Axx-Ayy}, this
introduces a correction of about 6\% to the values of $A_{yy}$.
These are shown in Fig.~\ref{Ayynp}, along with those for $A_y$,
for events with $E_{pn}<5{\ww}$MeV. Although the resolution in
momentum transfer was very good, the limited statistics produced
by the single telescope meant that the data were grouped in
$q$--bins of different size (20{\ww}MeV/c or 30{\ww}MeV/c). The
2.5{\ww}MeV threshold for protons in the telescope implies that
the lowest $q$--bin starts at 70{\ww}MeV/c and so the overlap with
the charge--exchange data set is not extensive. It is found that
$A_{yy}$ is negative but, in contrast to the charge--exchange
case, its magnitude decreases with increasing $q$. Furthermore,
whereas the charge--exchange $A_y$ was consistent with zero for
small $E_{pp}$, here, though small, it increases steadily in
magnitude with increasing $q$.

%
%

The impulse approximation description of the cross section and
analysing powers for the \dpce\ reaction was developed in
refs.~\cite{BW1,BW2,Carbonell} and we follow these works closely.
To show the basic sensitivity of our measurements, consider
neutron--proton charge--exchange amplitudes in the c.m.\ system:
\begin{equation}
f_{np}=\alpha +i\gamma
(\bmath{\sigma}_{n}+\bmath{\sigma}_{p})\cdot\hat{\bmath{n}} +\beta
(\bmath{\sigma}_{n}\!\cdot\hat{\bmath{n}})
(\bmath{\sigma}_{p}\!\cdot\hat{\bmath{n}})+
\delta(\bmath{\sigma}_{n}\!\cdot\hat{\bmath{m}})
(\bmath{\sigma}_{p}\!\cdot\hat{\bmath{m}})+\varepsilon
(\bmath{\sigma}_{n}\!\cdot\hat{\bmath{l}})
(\bmath{\sigma}_{p}\!\cdot\hat{\bmath{l}}),\label{fpn}
\end{equation}
where $\bmath{\sigma}_{n}$ and $\bmath{\sigma}_{p}$ are the Pauli
matrices for the neutron and proton. The basis vectors are defined
in terms of the initial ($\bmath{p}$) and final ($\bmath{p}'$)
momenta and lie along $\bmath{n}=\bmath{p}\times\bmath{p}'$,
$\bmath{l}=\bmath{p}'+\bmath{p}$, and
$\bmath{m}=\bmath{n}\times\bmath{l}$.


For small momentum transfers and low excitation energy $E_{pp}$ of
the final $pp$ pair, the \dpce\ charge exchange reaction mainly
excites the $^{1\!}S_0$ state of the final $pp$ system. The
spin--flip from the $pn$ triplet to $pp$ singlet provides a
\emph{spin--filter} mechanism. In single--scattering
approximation, the resulting amplitude depends upon the
spin--dependent parts of $f_{np}$, \emph{i.e.}\ $\beta$, $\gamma$,
$\delta$ and $\varepsilon$ but not the spin--independent term
$\alpha$. If, purely for the purposes of presentation, the
deuteron $D$--state is neglected, at low $E_{pp}$ we expect that
$A_{y}=0$ and
\begin{equation}
A_{xx}=\frac{|\beta|^2+|\gamma|^2+|\varepsilon|^2-2|\delta|^2}
{|\beta|^2+|\gamma|^2+|\delta|^2+|\varepsilon|^2}\:,\ \ \
A_{yy}=\frac{|\delta|^2+|\varepsilon|^2-2|\beta|^2-2|\gamma|^2}
{|\beta|^2+|\gamma|^2+|\delta|^2+|\varepsilon|^2}\:\cdot
\end{equation}

Since $\beta=\delta$ and $\gamma=0$ at $q=0$, the value of
$A_{xx}=A_{yy}$ depends there only on the ratio of $|\beta|$ to
$|\varepsilon|$, which is believed to change smoothly with
energy~\cite{BW2}. However, the $\delta$ amplitude, which contains
the one--pion--exchange pole, varies very rapidly with momentum
transfer and almost vanishes when $q\approx m_{\pi}$. Hence
$A_{xx}$ should approach its kinematical limit of $+1$ in this
region, and this is consistent with the trend of the data shown in
Fig.~\ref{Axx-Ayy}.

Spin--triplet final states generally produce opposite signs for
the analysing powers to the singlet states and so, away from the
very small $E_{pp}$ region, there is dilution of the $A_{xx}$ and
$A_{yy}$ signals. The estimation of this effect depends
sensitively upon the $pp$ final state interactions. In the program
of Ref.~\cite{Carbonell}, strong interactions~\cite{Sprung} were
kept only for $L\leq 2$, with the higher waves being distorted
solely by a point Coulomb force. The deuteron $S$-- and $D$--state
wave functions were taken from the Paris potential~\cite{Paris}
but it was verified that the use of more modern potentials for the
$pp$ and $np$ systems does not lead to any noticeable changes. The
predictions were made using amplitudes derived from the current
SAID $NN$ phase shift solution~\cite{SAID,Igor}.

It is known that, when $\vec{q}$ and the $pp$ relative momentum
$\vec{k}$ are perpendicular, odd partial waves cannot be excited
and the $pp$ system must be in a spin--singlet state~\cite{BW2}.
As a consequence, less triplet dilution of the analysing powers is
expected for small $\cos\theta_{qk}$. To show this, we have
divided the data shown in Fig.~\ref{Axx-Ayy} into the two regions
where $|\cos\theta_{qk}|\ {\small \lessgtr}\ 0.5$ and imposed the
same cuts on the theoretical description. All the features of both
$A_{xx}$ and $A_{yy}$ are then reproduced, including the variation
with $q$, $E_{pp}$, and $\cos\theta_{qk}$. It seems therefore that
the model is as valid here as at lower energies~\cite{Kox} and
that, as predicted in Ref.~\cite{BW2}, multiple scatterings do not
distort the analysing powers significantly.

Returning to the \dpnce\ reaction, the impulse approximation has
not been evaluated in such detail for these data so that only
qualitative statements can be made. If the deuteron $D$--state is
neglected, the transition from the $^3\!S_1$ deuteron bound state
to the $^3\!S_1$ $pn$ continuum is forbidden at $q=0$ due to the
orthogonality of the two wave functions. The only allowed
transition at $q=0$ is $^3\!S_1\to\, ^1\!S_0$, which has a
$(\Delta S,\,\Delta I,\,\Delta I_z)=(1,\,1,\,0)$ character. It is
the isobaric analogue of the deuteron charge--exchange amplitude
and should exhibit the same analysing powers. Furthermore, the
$(0,\,0,\,0)$ transitions, driven by the large isoscalar
spin--non--flip $NN$ amplitudes, vanish like $q^4$ at small $q$.
This is because they excite final $^3\!D_1$ or higher $S$--waves
that are orthogonal to the deuteron wave function. The behaviour
can be illustrated particularly clearly if one sums over all
excitation energies $E_{pn}$ to obtain the $\Delta S=0$ closure
sum rule~\cite{Franco,Dean}:
\begin{equation}
\left(\frac{\dd\sigma}{\dd t}\right)_{\!\!pd\to(pn)p}
=\half\left\{\left|\alpha_{pp}+\alpha_{pn}\right|^2
\left[1+S(q)-2S^2(\half q)\right]
+\left|\alpha_{pp}-\alpha_{pn}\right|^2
\left[1-S(q)\right]\right\}\:, \label{sumrule}
\end{equation}
where the $\alpha$ represent the spin--independent $NN$
amplitudes. Expanding the $S$--wave deuteron elastic form factor
$S(q)$ in powers of $q$, we see that the $\Delta I=1$ combination
$(\alpha_{pp}-\alpha_{pn})$ contributes to order $q^2$ whereas
cancellations for $\Delta I=0$ means that
$(\alpha_{pp}+\alpha_{pn})$ first contributes at order $q^4$.

To order $q^2$ the only source of dilution of the $A_{yy}$ signal
due to spin--triplet final states arises from $^3\!S_1\to\,
^3\!P_{0,1,2}$ transitions, which also involve an isospin flip.
Given that the final transition to this order, $^3\!S_1\to\,
^1\!P_{1}$, corresponds to spin--singlet final states, we would
expect that, for small momentum transfers, $A_{yy}$ should behave
much as for charge--exchange.

The simple picture must be modified to include the deuteron
$D$--state, though the basic suppression of the scalar--isoscalar
transitions at small $q$ remains. We therefore show in
Fig.~\ref{Ayynp} the predictions of the charge--exchange impulse
approximation for three upper limits on the value of $E_{pn}$.
Though at the lowest $q$ the $A_{yy}$ predictions are
qualitatively right, the analysing power decreases at higher $q$
while the predictions, even with a high 9{\ww}MeV cut--off, merely
level off. The charge--exchange program prediction for $A_y$
reaches barely 10\% of the \dpnce\ results shown in the figure,
though it must be stressed that vector analysing powers are
generally very sensitive to interference effects with small
amplitudes.

When the $\cos\theta_{qk}$ cut is applied, very little change in
the values of $A_{yy}$ is found in Fig.~\ref{Ayynp}. This very
different behaviour to that of charge exchange is due to the major
dilution effect here being engendered by the $I=0$, $L=0$ triplet
final state, which is clearly unaffected by the suppression of odd
partial waves by the angular selection. On the other hand, there
is a modest increase in $A_y$ for the larger $|\cos\theta_{qk}|$,
but any understanding of the significance of this will have to
await a more complete theoretical investigation.

In summary, we have measured the Cartesian analysing powers $A_y$,
$A_{xx}$ and $A_{yy}$ for the \dpce\ reaction at a beam energy of
$1170{\ww}$MeV, under kinematic conditions where both the momentum
transfer and the $pp$ excitation energy are small, and found good
agreement with the predictions of the impulse approximation. By
detecting other calibration reactions simultaneously in ANKE,
systematic effects arising from possible changes in luminosity
with polarisation state are avoided~\cite{Chiladze0}. The
implications of acceptance cuts must be studied in detail before
absolute cross sections can be obtained. Such information will be
vital for isolating the $\Delta I=0$ contribution by comparing the
\dpce\ and \dpnce\ cross sections.

The use of a single silicon telescope placed in the horizontal
plane meant that only $A_y$ and $A_{yy}$ could be determined for
the \dpnce\ reaction though in the future, with a system of four
larger telescopes~\cite{SPIN}, this limitation will be relaxed.
The results are far more complex than for charge exchange and a
more complete theoretical calculation is required to understand
them quantitatively. It is important in such a model that the
spin--triplet final states be treated consistently for the bound
and unbound systems, which means that the tensor force has to be
included also when evaluating the $pn$ continuum wave functions.

Having shown that the analysing powers in the charge--exchange
breakup reaction can be well described in impulse approximation,
these measurements will be extended to higher energies in order to
provide neutron--proton spin--dependent information in more barren
regions, and may be subsequently used as a polarimetry standard
for the COSY energy region. Although the maximum deuteron energy
available at COSY is only 1.15{\ww}GeV per nucleon, the energy
range can be extended to almost 3{\ww}GeV by inverting the
kinematics and using protons incident on a polarised deuterium
target~\cite{SPIN}. Furthermore, when both the proton and deuteron
are polarised, information will also be gained on the relative
phases of the $pn$ charge--exchange
amplitudes.\\

%
%
We are grateful to R.~Gebel, B.~Lorentz, H.~Rohdje\ss, and
D.~Prasuhn for the reliable operation of COSY and the deuteron
polarimeters. Neutron--proton scattering amplitudes corresponding
to the current SAID phase shifts were kindly provided by
I.I.~Strakovsky and we are also indebted to R.A.~Arndt for
discussions regarding the SAID observables.
%
%

\end{document}
%
%